\documentstyle[12pt]{article}
\begin{document}
\newcommand{\be}{\begin{eqnarray}}
\newcommand{\ee}{\end{eqnarray}}
\newcommand{\zt}{\zeta}
\newcommand{\ve}{\epsilon}
\newcommand{\al}{\alpha}
\newcommand{\gm}{\gamma}
\newcommand{\bt}{\beta}
\newcommand{\dt}{\delta}
\newcommand{\la}{\lambda}
\newcommand{\vp}{\varphi}
\newcommand{\nn}{\nonumber}
\newcommand{\I}{{\em I}}
\renewcommand{\baselinestretch}{1.2}
\begin{center}
{\bf On stability of the three-dimensional fixed point in a model with \\
three coupling constants from the $\ve$ expansion: Three-loop results.}
\end{center}
\vspace{0.5cm}
\begin{center}
 {\bf A. I. Mudrov}
\end{center}
\vspace{0.5cm}
\begin{center}
Department of Theoretical Physics,
Institute of Physics, St. Petersburg State University, Ulyanovskaya 1,
Stary Petergof, St. Petersburg, 198904, Russia\\
E-mail: aimudrov@dg2062.spb.edu
\vspace{0.5cm}
\begin{center}
 {\bf K. B. Varnashev}
\end{center}
\vspace{0.5cm}
Department of Physical Electronics, State Electrotechnical
University, \\
Prof. Popov Street  5, St. Petersburg, 197376, Russia
E-mail: root@post.etu.spb.ru
\end{center}
\vspace{1cm}

\begin{abstract}
The structure of the renormalization-group flows
in a model with three
quartic coupling constants is studied within the $\ve$-expansion
method up to three-loop order.
Twofold degeneracy of the eigenvalue
exponents for the three-dimensionally stable fixed point is observed
and the  possibility for powers in $\sqrt{\ve}$ to appear in the series
is investigated. Reliability and effectiveness of the
$\ve$-expansion method for the given model is discussed.
\end{abstract}
\vspace{3cm}


\vspace{1cm}

\newpage

\section{Introduction}
\label{sec:1}

In the this paper we consider the model given by the
effective Landau-Ginzburg-Wilson Hamiltonian with three quartic
coupling constants:
\begin{equation}
H =
\int d^Dx \Bigl[{1 \over 2}( m_0^2 \varphi_{\alpha}
\varphi_{\alpha}
 + \nabla \varphi_{\alpha}  \nabla \varphi_{\alpha})
+ {u_0 \over 4!} (\varphi_{\alpha} \varphi_{\alpha})^2
+\ {v_0 \over 4!} \varphi_{\alpha}^4
+ {2z_0 \over 4!} \varphi_{2\beta-1}^2 \varphi_{2\beta}^2
\Bigr].
\label{eq:1}
\end{equation}
Here $\vp_{\al}$ is a real vector order parameter field in
$D=4 - 2\ve$ dimensions, $\alpha = 1, 2,$ \ldots, $2N$,
$\beta = 1, 2,$ \ldots, $N$. The squared "bare mass" $m_0^2$ is a
linear measure of the temperature,
and $u_0$, $v_0$, and $z_0$ are the bare coupling constants.
Field $\vp_{\al}$ can be thought of as possessing two sets
of components, even and odd, each may be considered as an
$N$-component real vector.

The Hamiltonian (\ref{eq:1}) governs critical thermodynamics
in a number of interesting physical systems. For example,
when $N = 2$ it describes the structural phase transition
in $NbO_2$ crystal and antiferromagnetic phase transitions in $TbAu_2$
and $DyC_2$ for $v=z$. Another physically interesting case $N=3$ is
relevant to the antiferromagnetic phase transition in $K_2 Ir Cl_6$
crystal and, for $v = z$, to those in $Tb D_2$ and $Nd$ \cite{1,2}.
The detailed analysis of these systems along with the Landau
phenomenological theory can be found in Refs.\cite{1,3,4} with
references to experimental works therein.

For the first time the renormalization-group (RG) analysis of the
model (\ref{eq:1}) was carried out to the second order of
$\epsilon$ expansion  by Mukamel and Krinsky in Refs. \cite{1,2,3}. On
this ground, it was shown that the $2N$-component real anisotropic
model (\ref{eq:1}) possesses a unique (three-dimensionally) stable
fixed point for each $N \ge 2$.
On the other hand, the critical behavior
of this model  was studied within the field-theoretical
RG approach in three dimensions on the base of two- and three-loop
approximations \cite{5,6}.
There were obtained expansions for
$\bt$ functions and critical exponents for arbitrary $N$.
Using the generalized Pad$\acute{\rm e}$-Borel transformation,
the coordinates of the fixed points were found.
It was shown that a stable fixed point
did exist in the three-dimensional RG flow diagram for $N\ge 2$.

Assuming $v = z$,  model (\ref{eq:1})
formally turns into that with generalized cubic anisotropy and
a complex order parameter field.
The latter is a specific case ($m=2$) of the
well-known $mn$-component model. The critical thermodynamics of
this model was investigated in detail in Refs. \cite{7,8,9}.
Direct three-loop calculations for the case $m=2$, $n\ge2$
predict stability of the mixed
fixed point, the analog of the stable tetragonal fixed point
of model (\ref{eq:1}).

In the meantime, there are general arguments, not relying upon
perturbation theory, in favor of that a unique stable fixed point
should not be in the physical space although its existence is not
forbidden at $D>3$ \cite{10}. The same considerations lead to the
conclusion that the only three-dimensionally stable fixed point may
be  the Bose one and it is that point which governs the critical
thermodynamics in the phase transitions mentioned.
The point is that when $v=z$, model (\ref{eq:1}) describes $N$
interacting Bose systems. As was shown by Sak \cite{10a}, the
interaction term can be represented as the product of energy
operators of various two-component subsystems. It was also found
that one of the eigenvalue exponents characterizing the evolution
of this term under the renormalization group in the neighborhood
of the Bose fixed point is proportional to the specific-heat
exponent $\al$. Since $\al$ is believed to be negative at this point,
as confirmed by highly precise up-to-date experiments with liquid
helium \cite{11} including those in outer space \cite{12} and the
high-loop RG computations
carried out for the simple $O(n)$-symmetric  model in three
dimensions \cite{13,14}, the interaction is irrelevant. Therefore,
the Bose fixed point should be stable in three dimensions.

However, the RG approach, when directly applied to  model (\ref{eq:1})
and to the relative $mn$-component model, has not yet confirmed this
conclusion. On the contrary, all calculations performed
up to now indicate the existence of a unique stable
fixed point in the physical space, while the Bose point appears to be
three-dimensionally unstable [1--3,5--9]. This may be a consequence
of the  rather crude approximations used, and the higher order being taken
into account the closer the perturbative results could be to the
precise results. So, the aim of the paper is to investigate the critical
behavior of the three coupling constants model (\ref{eq:1}) in the
next, three-loop, order in $\ve$ and verify compatibility of the
predictions given by the $\ve$-expansion method with the other
techniques.

The main result of our study is that the unique fixed point, rather
than the Bose one, turns out to be three-dimensionally stable within
the given approximation. Calculation of the eigenvalue exponents of this
point is a nontrivial task due to their degeneracy in the one-loop
approximation. The analysis of the problem fulfilled in this paper
shows that such a degeneracy results in substantial reduction of the
information obtained from high-loop approximations.

\section{Three-loop $\bt$ functions, fixed points, and stability}
\label{sec:2}

The character of the critical asymptotics and the flow diagram
structure is known to be determined by the RG equations for quartic
coupling constants.
We calculate the perturbative expansions for the $\bt$-functions
for arbitrary $N$ within massless theory using dimensional
regularization and the minimal subtraction scheme. The  three-loop
results are as follows:
$$
\begin{array}{lll}
\bt_u &=& \ve u-u^2-\frac{1}{2 (N + 4)}\Bigl(6 u v + 2 u z \Bigr)+
\\[6pt]&&
\frac{1}{4 (N + 4)^2} \Bigl[12 u^3 (3 N + 7) + 132 u^2 v +
44 u^2 z + 30 u v^2 + 10 u z^2 \Bigr] -
\\[6pt]&&
\frac{1}{16 (N + 4)^3}
\Bigl[4 u^4 (48 \zt(3) (5 N + 11) + 33 N^2 + 461 N + 740) +
\\[6pt]&&
12 u^3 v (384 \zt(3) + 79 N + 659) +
4 u^3 z ( 384 \zt(3) + 79 N + 659) +
\\[6pt]&&
18 u^2 v^2 (96 \zt(3) + N + 321) + 1380 u^2 v z +
2 u^2 z^2 (288 \zt(3) + 3 N + \qquad\quad \quad \>
\\[6pt]&&
733) + 1512  u v^3 + 18 u v^2 z +
504 u v z^2 + 222 u z^3 \Bigr],\\[6pt]
\end{array}
$$
\begin{equation}
\begin{array}{lll}
\bt_v &=& \ve v-\frac{1}{2 (N + 4)} (12 u v + 9 v^2 + z^2) +
\\[6pt]&&
\frac{1}{4 (N + 4)^2} \Bigl[4 u^2 v (5 N + 41) + 276 u v^2 +
20 u v z + 24 u z^2 + 102 v^3 + 10 v z^2 +
\\[6pt]&&
8 z^3 \Bigr]-
\frac{1}{16 (N + 4)^3} \Bigl[8 u^3 v (96 \zt(3) (N + 7)-
13 N^2 + 184 N + 821) +
\\[6pt]&&
18 u^2 v^2 (768 \zt(3) + 17 N + 975) +
12 u^2 v z (96 \zt(3) - 13 N + 154) +
\\[6pt]&&
2 u^2 z^2 (576 \zt(3) + 43 N + 667) +
108 u v^3 (96 \zt(3) + 131) + 306 u v^2 z +
\\[6pt]&&
12 u v z^2 (96 \zt(3) + 187) +
2 u z^3 (384 \zt(3) + 395) +
27 v^4 (96 \zt(3) + 145) +
\\[6pt]&&
162 v^2 z^2 +
8 v z^3 (48 \zt(3) + 101) +
3 z^4 (32 \zt(3) + 17) \Bigr] ,\\[6pt]
\end{array}
\label{eq:2.6}
\end{equation}
$$
\begin{array}{lll}
\bt_z &=& \ve z-\frac{1}{2 (N + 4)} (12 u z + 6 v z + 4 z^2) +
\\[6pt]&&
\frac{1}{4 (N + 4)^2} \Bigl[4 u^2 z (5 N + 41) + 204 u v z +
116 u z^2 + 30 v^2 z + 72 v z^2 + 18 z^3 \Bigr]-
\\[6pt]&&
\frac{1}{16 (N + 4)^3} \Bigl[8 u^3 z (96 \zt(3) (N + 7)
-13 N^2 + 184 N + 821) +
\nn\\[6pt]&&
12 u^2 v z (864 \zt(3) + 4 N + 1129) +
4 u^2 z^2 (1440 \zt(3) + 47 N + 1796) +
\\[6pt]&&
18 u v^2 z (192 \zt(3) + 391) + 72 u v z^2 (96 \zt(3) + 103) +
2 u z^3 (960 \zt(3) + 1517) +
\\[6pt]&&
1512 v^3 z +
36 v^2 z^2 (48 \zt(3) + 35) +
72 v z^3 (16 \zt(3) + 25) +
4 z^4 (48 \zt(3) + 91) \Bigr],
\end{array}
$$
where $\zt$ is the Riemann $\zt$ function: $\zt(3)=1.20206$.
The model under consideration is known to have eight fixed
points \cite{2,5}. Below we write out  the coordinates of the
most interesting II-tetragonal fixed point only.
\begin{equation}
\begin{array}{lll}
\\[6pt]
 u_c &=&
 \frac{N+4}{(5N-4)}\ve+
\frac{N+4}{(4-5 N)^3}(70 N^2-205 N+139)\ve^2+
\\[6pt]&&
 \Bigl(\frac{12(N+4)}{(5 N-4)^4}\zt(3)(64N^3-188N^2+151N-23)+
\\[6pt]&&
 \frac{N+4}{4(4-5N)^5}(6370N^4+24149N^3-144719N^2+197208N
 -83256)\Bigr)\ve^3
,\\[6pt]
 v_c &=&
 \frac{N+4}{(5 N-4)}(N-2)\ve+
\frac{N+4}{(5 N-4)^3}(30 N^3+25 N^2-217 N+166)\ve^2-
\\[6pt]&&
 \Bigl(\frac{24(N+4)}{(5 N-4)^4} \zt(3)(8N^4+16N^3-88N^2+75N-9)
 -\frac{N+4}{4 (5 N-4)^5}(1030 N^5+
\\[6pt]&&
 2751 N^4+46033 N^3-207590 N^2+267336 N-109808)
\Bigr)\ve^3
,\\[6pt]
 z_c &=&
 \frac{N+4}{5 N-4} (N-2) \ve+
\frac{N+4}{(5 N-4)^3}(30 N^3+25 N^2-217 N+166) \ve^2-
\\[6pt]&&
 \Bigl(\frac{24(N+4)}{(5 N-4)^4}\zt(3) (8 N^4+16 N^3-88 N^2+75 N-9)-
\frac{N+4}{4 (5 N-4)^5} (1030 N^5+
\\[6pt]&&
 2751 N^4+46033 N^3-207590 N^2+267336 N-109808)\Bigr)\ve^3
.\\[6pt]&&
\end{array}\label{coordinates}
\end{equation}
In order to determine the character of stability of this point
we should calculate the stability matrix eigenvalues $\la$'s.
It is convenient, rather, to deal with the quantity
$y={\la\over \ve}$ being a root of the reduced
characteristic polynomial hereafter denoted  $P(y,\ve)$:
\be
 - y^3(\ve) + a(\ve) y^2(\ve) -  b(\ve) y(\ve) + c(\ve) &=& 0.
\label{eq:B.1}
\ee
The coefficients  $a(\ve)$, $b(\ve)$, and  $c(\ve)$ are the  formal series
\be
a(\ve)&=&a_0+a_1 \ve+a_2 \ve^2+\ldots ,\nn\\
b(\ve)&=&b_0+b_1 \ve+b_2 \ve^2+\ldots , \label{eq:B.2}\\
c(\ve)&=&c_0+c_1 \ve+c_2 \ve^2+\ldots .\nn
\ee
A solution $y(\ve)$ to Eq. (\ref{eq:B.1}) is obtained by
consecutive calculating  coefficients of the series
$y(\ve)=\sum y_k \ve^k $
in corresponding orders in $\ve$.
Normally, the polynomial $P(y,0)$ has three different roots
$y_0$, and the derivative $\partial_y P(y_0,0)$ does not vanish. It
implies that the coefficients $y_k $ are determined in the  $k$-th
order in $\ve$. This customary scheme does not work when the
polynomial $P(y,0)$ has multiple roots. It takes place for the
II-tetragonal fixed point of model (\ref{eq:1}) since the
one-loop approximation, as will be shown below (see formula
(\ref{y_0})), yields two equal values of $y_0$. To treat this situation
properly we have to  thoroughly analyze the problem of expanding
such twofold degenerate solutions.

If coefficients $a(\ve)$, $b(\ve)$, and  $c(\ve)$ themselves
had been
polynomials, the solution $y(\ve)$ would have belonged to the class
of so-called algebraic functions. Such a function is analytical
on the complex plane, except for a finite set of isolated points,
where it has either poles or branchings of a finite order. The
poles are associated with the zeros of the highest coefficient
in $P(y,\ve)$, while the branching points are associated with those
values of
$\ve$ where $P(y,\ve)$ has multiple roots. With twofold
degeneracy of $y_0$, branching of order $2$ is possible  at
the point $\ve=0$. But a two-valued function cannot be expanded
 in integer powers of $\ve$. Instead, it should be
represented by a Puiseux series in powers of $\sqrt\ve$  \cite{15}.
Similar speculations are valid when the dependence of $P(y,\ve)$
on $\ve$ is of the formal series type since every coefficient of
the expansion $y(\ve)$ is determined by a finite number of terms
in $a(\ve)$, $b(\ve)$, and  $c(\ve)$. Therefore, whatever the
perturbative order is chosen, there is an algebraic function
coinciding with $y(\ve)$ modulo higher terms. So, let us formulate
the anzatz for $y(\ve)$ as
\be
y(\ve)&=&y_0+y_{1\over 2} \ve^{1 \over 2}+y_1 \ve+y_{3\over 2}
\ve^{3 \over 2}+\ldots .
\label{eq:B.3}
\ee
We shall show that a solution to Eq. (\ref{eq:B.1}) in the case
of twofold degenerate roots at $\ve=0$ does exist in form
(\ref{eq:B.3}). We are also interested when noninteger powers
in $\ve$ do appear in and when they  drop from $y(\ve)$. The answer
will be given by a theorem before which we introduce some notations.
Let ${\bf Z^+/2}$ be the set of non-negative half-integer numbers.
We assume that the infinite point $\infty$ also belongs to
${\bf Z^+/2}$. It is convenient to think of it as of an integer
number. Denote $[a,b]$ the interval in ${\bf Z^+/2}$ with
boundaries $a$ and $b$, i.e., the set of points
$l \in {\bf Z^+/2}$ satisfying inequality $a\leq l\leq b $.
To distinguish intervals without one or two boundary points
we use parentheses instead of square brackets.\\[12pt]
{\bf Theorem. }{\em In the case of twofold degenerate roots
in the one-loop approximation, the solutions $y(\ve)$ to Eq.
(\ref{eq:B.1}) are represented by series in powers of $\sqrt{\ve}$.
There is an alternative: either $P(y,\ve)$ has two equal roots
in every order of the perturbation theory or the solution  $y(\ve)$
splits at a finite step $l_s$. Noninteger powers of $\ve$
contribute to the expansion $y(\ve)$ if and only if $l_s$ is
a noninteger number.
}\\[12pt]
Let us define polynomials
$A_l(y_0,y_{1 \over 2},...,y_l)$,  $l \in {\bf Z^+/2}$,
from the expansion
$$\frac{\partial P(y,\ve)}{\partial y}=\sum_{l \in {\bf Z^+/2}}
          A_l(y_0,y_{1 \over 2},...,y_l) \ve^l.$$
Another way to introduce them is as follows.
Consider the $\ve$ expansion of the reduced characteristic
equation (ERCE) obtained by substitution of Eqs. (\ref{eq:B.2}) and
 (\ref{eq:B.3})
into Eq. (\ref{eq:B.1}). At a sufficiently high order the
coefficient before $\ve^m$ may be shown to be represented as a sum
$$ A_0(y_0) y_m +  A_{1 \over 2}(y_0,y_{1 \over 2}) y_{m-{1 \over 2}}
+\ldots+
A_l(y_0,y_{1 \over 2},...,y_l) y_{m-l} + \ldots =0,\quad m-l>l.$$
Further, let $\I=[0,l_s]$ be the maximum interval containing $0$
such that for $l\in \I$ the coefficient $y_l$ is found from
the order $2l$ of ERCE.
It implies, in particular, that the partial solution
$[y]_l\equiv (y_0,y_{1\over 2},...y_l)$ to Eq. (\ref{eq:B.1})
exists for all $l\in [0,l_s)$.
The upper boundary $l_s$ may be  either finite or infinite.
First let us prove that the polynomials
$A_l(y_0,y_{1 \over 2},...,y_l)$ for $l\in [0,l_s)$ turn zero
upon substitution of the partial solution $[y]_l$ into them.
This is the case, at least, for $l=0$. Supposing it
for all $l$  such that $0\leq l< m$, where $m$ is some
half-integer number strictly below $l_s$, we have
$$[ A_0(y_0) y_{2m+{1 \over 2}} + A_{1 \over 2}(y_0,y_{1 \over 2})
y_{2m} +\ldots]+
A_m(y_0,y_{1 \over 2},...,y_m) y_{m+{1 \over 2}} +\ldots=0$$
in the order $2m+{1 \over 2}$ of ERCE.
The expression within the square brackets vanishes due to the
assumption made, while  the  rightmost dots  stand for the terms
depending on $y_l$ with $l$ less than ${m+{1 \over 2}}$.
Thus, if $A_m(y_0,y_{1 \over 2},...,y_m)\not = 0$, the
coefficient $y_{m+{1 \over 2}}$ is determined from the order
$2m+{1 \over 2}\not= 2m+1$, in contradiction to
$ m+{1 \over 2}\leq l_s$ and  $m+{1 \over 2} \in \I$.
The consequence of the fact just stated is that all
$y_l$ with $l\in [0,l_s)$ are found from a quadratic equation
$$Q_l(y_l)\equiv {1\over 2}\partial^2_y P(y_0,0) y^2_l+
L_l(y_0,y_{1 \over 2},...,y_{l-{1 \over 2}})y_l+
R_l(y_0,y_{1 \over 2},...,y_{l-{1 \over 2}})=0.$$
The highest coefficient  ${1\over 2}\partial^2_y P(y_0,0)$ is
nonzero as only two of the three roots $y_0$ coincide.
Now we can give one more characteristic to the polynomial
$A_l(y_0,y_{1 \over 2},...,y_l)$ for $l\in \I$.
Namely, it is the  derivative of $Q_l(y_l)$
with respect to $y_l$. It follows directly from here that
the solution $y(\ve)$ does not split at the orders
$l\in[0,l_s)$, that is every polynomial $Q_l(y_l)$ has equal
roots. Another implication is that all noninteger
$l\in[0,l_s)$ give $y_l=0$. Indeed, every term of the
linear coefficient
$L_l(y_0,y_{1 \over 2},...,y_{l-{1 \over 2}})$ of  $Q_l(y_l)$
depends on some variable $y_k$ with noninteger numbers  $k<l$
(because none of the series $a(\ve)$, $b(\ve)$, and  $c(\ve)$
involves noninteger powers in $\ve$) and vanishes once they
turn zero. Assuming recurrently $y_k=0$ for noninteger $k<l$,
we find  $y_l$ obeying  the equation
$Q_l(y_l)={1\over 2}\partial^2_y P(y_0,0) y^2_l+
R_l(y_{1 \over 2},...,y_{l-{1 \over 2}})=0$.
Since $Q_l(y_l)$ has two equal roots, both are zero.
We shall show now that with finite $l_s$, the solution
$y(\ve)$ splits at the order $l_s$ or, equivalently,
that  $A_{l_s}(y_0,y_{1 \over 2},...,y_{l_s})\not = 0$.
Supposing the opposite, consider the coefficient of ERCE
in the noninteger order $2l_s+{1\over 2}$ in $\ve$:
$$[A_0(y_0) y_{2l_s+{1\over 2}}
+A_{1 \over 2}(y_0,y_{1 \over 2})y_{2l_s}
+\ldots+ A_{l_s}(y_0,y_{1 \over 2},...,y_{l_s})y_{l_s
+{1\over 2}}]+\ldots=0.$$
The expression within the square brackets vanishes as well as
the terms depicted by the dots on the right.
Those terms contain $y_l$ with noninteger numbers $l$ from the
interval $\I$. As was stated above, such $y_l$ are equal to zero,
hence in the order  $2l_s+{1\over 2}$ Eq. (\ref{eq:B.1}) holds
identically. The next order $2l_s+1$ gives a quadratic equation
for $y_{l_s+{1\over 2}}$ with the highest coefficient
${1\over 2}\partial^2_y P(y_0,0)\not = 0$.
It means that  $l_s+{1\over 2}$ belongs to the interval $\I$ that
contradicts the assumption about $l_s$ being its
upper boundary. So, $A_{l_s}(y_0,y_{1 \over 2},...,y_{l_s})\not = 0$
and for arbitrary $m \in {\bf Z^+/2}$ the quantity $y_{l_s+m}$
is determined in the order $2l_s+m$ in $\ve$ from a linear equation
in which $A_{l_s}(y_0,y_{1 \over 2},...,y_{l_s})$ is the highest
coefficient. Thus we have proved the existence of solutions to the
reduced characteristic equation (\ref{eq:B.1}) in the form
(\ref{eq:B.3}). We have yet to verify disappearance of  noninteger
powers of $\ve$ from expansion $y(\ve)$ provided $l_s$ is an integer
number. As was shown, the coefficient $y_{l_s+m}$  is determined in
the order $2l_s+m$ from a linear equation of the form
$A_{l_s}(y_0,y_{1 \over 2},...,y_{l_s}) y_{l_s+m}+
B_{l_s+m}(y_0,y_{1 \over 2},...,y_{l_s+m-{1 \over 2}})=0$.
Supposing $l_s+m$ and therefore $m$ to be noninteger, we
see that each term in $B_{l_s+m}$ depends on some $y_l$ with
noninteger number $l<l_s+m$ (because none of $a(\ve)$, $b(\ve)$,
and  $c(\ve)$ contains powers of $\sqrt \ve$). It has already been
shown that $y_l=0$  for noninteger $l$ from the interval $\I$. Assuming
recursively $y_l=0$ for $l<l_s+m$, we have
$B_{l_s+m}(y_0,y_{1 \over 2},...,y_{m-{1 \over 2}})=0$
and therefore $y_l=0$ for all noninteger $l$'s.

Let us apply  the theorem proved to the problem of calculating
eigenvalue exponents for the II-tetragonal fixed point.
The explicit form of the coefficients in Eq. (\ref{eq:B.2}) reads
\begin{equation}
\begin{array}{lll}
a_0&=&\frac{-1}{(5N-4)}(7 N - 8)    , \nn\\
a_1&=&\frac{1}{(5N-4)^3}(270 N^3 - 1129 N^2 + 1591 N - 736 ) ,\nn\\
a_2&=&\frac{-1}{2 (5N-4)^5}(48 \zt(3)(5 N - 4)(144 N^4 - 720 N^3 +
1289 N^2 - 947 N +230) \nn\\
   &+& 10030 N^5 - 104229 N^4 + 429747 N^3 - 804632 N^2 + 691620 N
   - 222720) ,\nn\\
b_0&=&\frac{1}{(5N - 4)^2}(N - 2)(11 N - 10)    ,\nn\\
b_1&=&\frac{-2}{(5N - 4)^4}  (510 N^4 - 3157 N^3+6615 N^2 -
5832 N + 1868)     ,\nn\\
b_2&=&\frac{1}{(5N - 4)^6}(48 \zt(3)(5 N - 4) (272 N^5 - 1824 N^4 +
4455 N^3 - 5095 N^2 \nn\\
   &+& 2754 N -558) +  25890 N^6 - 338437 N^5 + 1547050 N^4 -
3437182 N^3 \nn\\
   &+& 4044203 N^2  - 2430752 N + 589412)     
\end{array}\label{eq:B.4}
\end{equation}
$$
\begin{array}{lll}
c_0&=&\frac{-1}{(5N - 4)^2}(N - 2)^2     ,\nn\\
c_1&=&\frac{1}{(5N - 4)^4}(N - 2)(150 N^3 - 809 N^2 +
1229 N - 566)     ,\nn\\
c_2&=&\frac{-1}{2 (5N - 4)^6}(48 \zt(3)(N - 2)(5 N - 4)(80 N^4
- 464 N^3 + 865 N^2 \nn\\
   &-&  641 N + 164)  + 13950 N^6 - 184745 N^5 + 887705 N^4 -
2072060  N^3 \qquad\qquad\nn\\
   &+&     2541094  N^2 - 1575640 N +389512)           .\nn
\end{array}\nn
$$
Substituting this  into  Eq. (\ref{eq:B.1}) and setting $\ve=0$
(one-loop approximation) we find $y_0$ to be twofold degenerate:
\begin{equation}
y^{(1)}_0=-1,\quad y^{(2)}_0=y^{(3)}_0=\frac{2-N}{5N-4}.\label{y_0}
\end{equation}
The solution $y(\ve)$ expanding simple root $y^{(1)}_0$ is
calculated in the conventional way. Our further consideration
concerns the multiple root only. The first appearance of the
coefficient $y_{1\over 2}$ occurs in the order $1\over 2$ in
$\ve$ of ERCE, with multiplier $A_0(y_0)=\partial_y P(y_0,0)$:
                            $$y_{1\over 2}A_0(y_0)=0.$$
Due to the degeneracy of $y_0$, $A_0(y_0)=0$ and $y_{1\over 2}$
cannot be  actually determined from the  order ${1\over 2}$.
To find $y_{1\over 2}$ we must solve the quadratic equation
in the order 1 of ERCE:
\be
Q_{1\over 2}(y_{1\over 2})&=& {1\over 2}\partial^2_y P(y_0,0)
y^2_{1\over 2} + (a_1 y_0^2 - b_1 y_0+c_1)=0.\label{eq:B.5}
\ee
The highest coefficient
${1\over 2}\partial^2_y P(y_0,0)=-(3 y_0 - a_0)$
is nonzero because only two of the three roots $y_0$ coincide.
Substitution  of Eq. (\ref{eq:B.4}) into  Eq. (\ref{eq:B.5}) gives
$y_{1\over 2}=0$ for all $N$. The next order ${3\over 2}$ of
ERCE does not provide $y_1$ because at this step the equation holds
identically:
$$A_0(y_0)y_{3\over 2}+ A_{1\over 2}(y_0,y_{1\over 2})y_1+
(...)y_{1\over 2}=0.$$
Here $A_{1\over 2}(y_0,y_{1\over 2})=
\partial_{y_{1\over 2}}Q_{1\over 2}(y_{1\over 2})=
-2(3 y_0 - a_0)y_{1\over 2}=0$.
Considering the factor before $\ve^2$ in ERCE, we come to the quadratic
equation
\be
  Q_1(y_1)&=& {1\over 2}\partial^2_y P(y_0,0) y^2_1 +
(2 a_1 y_0-b_1) y_1 + a_2 y^2_0- b_2 y_0 + c_2=0, \label{eq:B.6}
\ee
which has the solution
\be
 y_1&=&\frac{3(N-1)(40N^3-208N^2+253N-66)}{(2N-1)(5N-4)^3} \nn\\
    &\pm& \frac{4 |(N-1)(N-2)(N+4)(5N-4)|}{(2N-1)(5N-4)^3}.
     \label{eq:B.7}
\ee
The order ${5\over 2}$ of ERCE gives rise to the relation
$$ [A_0(y_0)y_{5\over 2}+  A_{1\over 2}(y_0,y_{1\over 2})y_2]+
   A_1(y_0,y_{1\over 2},y_1)y_{3\over 2}+(...)y_{1\over 2} =0,$$
where the expression within the square brackets is zero and
\be
  A_1(y_0,y_{1\over 2},y_1)=-6 y_0 y_1+2 a_0 y_1+ 2 a_1 y_0 - b_1.
\label{eq:B.8}
\ee
As was shown when proving the theorem, $A_1(y_0,y_{1\over 2},y_1)$ is
the derivative of the quadratic polynomial $Q_1(y_1)$ with respect
to $y_1$, formula (\ref{eq:B.6}). Hence, $A_1(y_0,y_{1\over 2},y_1)$
vanishes if and only if $Q_1(y_1)$ has two equal roots.
It occurs when $N=1$ and $N=2$ in formula (\ref{eq:B.7}).

Thus, for $N\not =1,2$ the solution $y(\ve)$ splits at the integer
order in $\ve$ and, according to the theorem, would not contain
noninteger powers of $\ve$. We cannot make such a statement for
the physically significant case $N=2$ because the coefficient
$y_{3\over 2}$ is determined in the third order in $\ve$ of ERCE
involving four-loop contributions. If $y_{3\over 2}$ comes out to
be nonzero, the expansion $y(\ve)$ will contain noninteger
powers. Otherwise, $y_{3\over 2}=0$ and only the five-loop
approximation, as follows from the theorem, will allow us to
calculate the next coefficient $y_2$ in Eq. (\ref{eq:B.3}).

To sum up, the three-loop eigenvalue exponents for
the II-tetragonal fixed point read
\begin{equation}
\begin{array}{lll}
&&\\[6pt]
 \la_1 &=& -\ve+\frac{1}{(5 N-4)^2 (2 N-1)}(60 N^3-160 N^2+181 N-85)
 \ve^2+
 \\[6pt]&&
\frac{1}{2 (5 N-4)^4 (1-2 N)^3}\Bigl(48 \zt(3) (2 N-1)^2 (5 N-4)
(32 N^4-128 N^3+212 N^2-
 \\[6pt]&&
153 N+33)+20560 N^7-165328 N^6+644392 N^5-1406864 N^4+
 \\[6pt]&&
1756745 N^3-1224341 N^2+433704 N-59052\Bigr)
\ve^3
,\\[6pt]
 \la_2 &=& \frac{2-N}{5 N-4}\ve+\frac{1-N}{(5 N-4)^3 (2 N-1)}
\Bigl(4 \> sgn(N-1) |5 N^3+6 N^2-48 N+32|-
\\[6pt]&&
3 (40 N^3-208 N^2+253 N-66)\Bigr) \ve^2
,\\[6pt]
 \la_3 &=& \frac{2-N}{5 N-4}\ve+\frac{(N-1)}{(5 N-4)^3 (2 N-1)}
\Bigl(4 \> sgn(N-1) |5 N^3+6 N^2-48 N+32|+
 \\[6pt]&&
3 (40 N^3-208 N^2+253 N-66)\Bigr)\ve^2
.\\[12pt]
\end{array}
\end{equation}
For the physically interesting cases  $N=2$ and $N=3$ they are
presented in Table I.
The eigenvalue exponents for the Bose fixed point, the calculation of
which is an easy task, are also written out for comparison.
\begin{table}
\caption{Three-loop eigenvalue exponents for the Bose and the
II-tetragonal fixed points ($\ve={1\over 2}$ corresponds to
the physical space)}
\vspace{0.5cm}
\hspace{1cm}
\begin{tabular}{|l|l|l|}\hline
Type      &     &
  \\
of fixed      &\multicolumn{1}{|c|}{$N=2$}& \multicolumn{1}{|c|}{$N=3$}
  \\
point
& &
  \\      \hline
& \multicolumn{2}{|c|}{}
  \\       &\multicolumn{2}{|c|}{
            $\la_u= {1 \over 5}\ve -{14 \over 25}\ve^2
            +\frac{768\zt(3)-311}{625}\ve^3$}
  \\[6pt]
 Bose      &\multicolumn{2}{|c|}{
            $\la_1=-{1 \over 5}\ve +{2 \over 5}\ve^2
            -\frac{768 \zt(3)+29}{625}\ve^3$}
  \\[6pt]  &\multicolumn{2}{|c|}{
            $\la_2=- \ve +{6 \over 5}\ve^2
            -\frac{384\zt(3)+257}{125}\ve^3$}
  \\[8pt] \hline
& &
  \\
II-tet-
           &$\la_1=-\ve +{13 \over 12}\ve^2
           -\frac{84\zt(3)+65}{36}\ve^3$
           &$\la_1=-\ve +{58\over 55 }\ve^2
           -\frac{3(123600\zt(3)+71621)}{166375}\ve^3$
  \\[6pt]
ragonal
           &$\la_2=-{1 \over 3}\ve^2$
           &$\la_2=-{1 \over 11}\ve - {2 \over 11}\ve^2$
  \\[6pt]
           &$\la_3=-{1 \over 3}\ve^2$
           &$\la_3=-{1 \over 11}\ve +{2 \over 605}\ve^2$
  \\[8pt] \hline
\end{tabular}
\end{table}
It follows from the table that the II-tetragonal fixed point is
absolutely stable in three dimensions ($3D$), in contrast to the Bose one.
Obviously, simple resummation procedures, such as
the Pad$\acute{\rm e}$ and Pad$\acute{\rm e}$-Borel methods,
being applied to $\la$'s,  do not change the picture.

Let us now calculate the critical dimensionality $N_c$ of the order
parameter. Its $\ve$ expansion is found from condition $v_c=z_c=0$
imposed on the right-hand side of Eq. (\ref{coordinates}):
\be
 N_c&=& 2 - 2 \ve + {5\over 6}(6 \zt(3)-1) \ve^2+ O(\ve^3).
 \label{N_Ó}
\ee
The critical dimensionality separates two different regimes of
critical behavior of the model. For $N>N_c$ the II-tetragonal
rather than the Bose fixed point is three-dimensionally stable
in $3D$. At $N=N_c$ they interchange their stability so that for
$N<N_c$ the stable fixed point is the Bose one.

Because the series (\ref{N_Ó}) is alternating, it can be resummed
by means of the Pad$\acute {\rm e}$-Borel method, the result being
\be
 N_c&=& a - {2b^2\over c} + {4b^3\over {c^2\ve}}
 \exp\Biggl(- {2b\over {c\ve}}\Biggr)
 E_i\Biggl({2b\over {c\ve}}\Biggr).
 \label{N_c_Resum}
\ee
Here $a$, $b$, and $c$ are the coefficients before $\ve^0$, $\ve^1$
and $\ve^2$ in Eq. (\ref{N_Ó}), respectively, and  $E_i(x)$ is the
exponential integral. Setting $\ve={1\over 2}$, from
Eq. (\ref{N_c_Resum}) we obtain  the value of critical dimensionality
\be
 N_c&=& 1.50 \>.
 \label{eq:4.3}
\ee
Since $N_c$ lies below two, within the given approximation the critical
behavior of model (\ref{eq:1}) for $N=2$ and $N=3$ must be
governed by the II-tetragonal fixed point. It confirms the
deductions given by the analysis of the eigenvalue exponents.

\section{Discussions}
\label{sec:3}

In conclusion, let us briefly discuss the results of the present
investigation.
The structure of the RG flows of the three quartic coupling constants
model was studied within the $\ve$-expansion method.
The three-loop series for $\bt$ functions of
the model were obtained.
Eigenvalue exponents for the most intriguing II-tetragonal
and the Bose fixed points were calculated for arbitrary $N$.
For the physically interesting values $N=2$ and $N=3$, the
II-tetragonal rather than the Bose fixed point
was shown to be three-dimensionally stable in $3D$,
and the critical dimensionality $N_c=1.5$ found has confirmed
this conclusion. Consequently, the critical thermodynamics of the
antiferromagnetic phase transitions in such substances as $TbAu_2$,
$DyC_2$, $K_2 Ir Cl_6$, $Tb D_2$, and $Nd$ as well as the structural
phase transition in $NbO_2$ crystal should
be controlled by the II-tetragonal fixed point.
It agrees with the results by  Mukamel and Krinsky \cite{1,2,3} as
well as with calculations performed within the field-theoretical RG
approach in $3D$ (Refs. \cite{5,6}) but contradicts to the  non-perturbative
inferences \cite{10}.
The distinction of $3D$ RG  predictions from those of the precise
theory may be regarded as an effect of low-order approximations.
The point is that the three-loop analysis in $3D$ (Refs. \cite{5,6})
shows the two rival fixed points (Bose and II-tetragonal) to be
close to one another,  and it is natural to expect that taking
into account next perturbative terms may change their character of
stability properly. Such speculations, however, do not suit the
$\ve$-expansion because we can judge about the closeness of points just
indirectly, i.e. by comparing their critical exponents.
The $\ve$-expansion analysis yields the critical exponents of the
II-tetragonal fixed point to be considerably distant from those of
the Bose fixed point \cite{1,2,6,16}, and hardly one can hope  that
longer $\ve$ series even resummed will bring them close to one
another.
All this allows us to raise the question whether the
$\ve$-expansion method is reliable for the given model.

The twofold degeneracy of the solutions to the characteristic
equation for the II-tetragonal fixed point in the one-loop
approximation worsens the situation. According to the analysis
performed, the eigenvalue exponents should be represented as
series in powers of $\sqrt{\ve}$ instead of $\ve$.
Even if we adopt the idea that noninteger powers  actually drop from the
expansions, such a degeneracy decreases the accuracy expected
within a given approximation.
Namely, in the frame of three-loop approximation we effectively
obtain two-loop-like pieces of the  series, and to evaluate the
next term (of order $\ve^3$) we must take into account the
five-loop contributions \cite{17}.
So, computational difficulties grow faster than
the amount of essential information one may extract from the
high-loop approximations.
This leads to the conclusion that the $\ve$-expansion method is
not quite effective for the given model.

Of course, the $\ve$-expansion method is perfectly reliable in the
four-dimensional space time, i.e., in studying field systems.
However, as the present investigation shows, one should be careful
when applying it to complicated three-dimensional models of
statistical physics, especially
if insufficiently high approximations are used.

\noindent
{\bf Acknowledgments}\\[12pt]

We thank Prof. A. I. Sokolov for useful discussions. One of us
(K.B.V.) appreciates the support by the Russian Research Program
"Fullerens and Atomic Clusters" (Grant No. 94024).


\begin{thebibliography}{99}
\bibitem{17} Similar phenomenon was observed
earlier in the course of studying  the impure Ising model  (see Ref. [19] and
references therein).
Half-integer powers in $\ve$ arising in that model have different
origin but also lead to the loss of accuracy.
\end{thebibliography}
\end{document}